\documentclass[a4paper,11pt]{article}
\usepackage{pos}
\usepackage{hyperref}
\usepackage{float}
\usepackage{placeins}
\newcommand{\doilink}[2]{\href{https://doi.org/#2}{#1}} 
\newcommand{\newlink}[2]{\href{#2}{#1}}

\title{Modeling of Dark Matter Prompt and Secondary Signatures in Dwarf Galaxies}

\author*[a,b]{Athithya Aravinthan}
\author[a,b]{Julien Dörner}
\author[a,b,c]{Julia Becker Tjus}
\author[d,e]{Aritra Basu}
\author[a,b]{Dominik Bomans}
\author[a,b]{Samata Das}
\author[a,b, h]{Sam Taziaux}
\author[b,f]{Dominik Elsässer}
\author[g]{Riccardo Catena}

\affiliation[a]{Ruhr University Bochum, Faculty of Physics and Astronomy, Plasma-Astroparticle Physics, 44780 Bochum, Germany}
\affiliation[b]{Ruhr Astroparticle and Plasma Physics Center (RAPP Center), Germany}
\affiliation[c]{Department of Space, Earth and Environment, Chalmers University of Technology, SE-412 96 Gothenburg, Sweden}
\affiliation[d]{Thuringian State Observatory (TLS), Sternwarte 5, 07778 Tautenburg, Germany}
\affiliation[e]{Max Planck Institute for Radio Astronomy, Auf dem Hügel 69, 53121 Bonn, Germany}
\affiliation[f]{TU Dortmund University, Department of Physics, 44221 Dortmund, Germany}
\affiliation[g]{Department of Physics, Chalmers University of Technology, SE-412 96 Gothenburg, Sweden}
\affiliation[h]{CSIRO Space and Astronomy, PO Box 1130, Bentley WA 6102, Australia}

\emailAdd{athithya.aravinthan@rub.de}

\abstract{Dwarf Spheroidal (dSph) galaxies are very promising laboratories for the indirect search for dark matter (DM), due to their low astrophysical background in radio and gamma-ray frequencies. This is convenient when considering Weakly Interacting Dark Matter (WIMP) that can annihilate and produce radio continuum emission. 

Radio detections of dSph galaxies, however, prove to be difficult and motivate the consideration of transient galaxies that have just recently become quiescent.  

For the past several decades, the prompt emission from DM annihilation signatures has been explored through modeling and the setting of limits. In addition to the prompt annihilation signatures from neutrinos, gamma-rays, electrons, positrons, and antimatter, the secondary emission from charged annihilation products undergoing radiative loss processes also contributes to the picture. For instance, synchrotron radiation and inverse Compton scattering of charged products such as electrons and positrons can provide a significant signal. The quantitative modeling of this secondary emission with the astrophysical background is necessary to place stringent constraints on the nature of DM.

In this work, the multi-wavelength secondary spectrum of DM annihilation for a dwarf galaxy is calculated using the open-source code \texttt{CRPropa 3.2}, which enables the self-consistent treatment of the astrophysical background and secondary emissions. We present a systematic comparison of signatures from conventional astrophysical processes to those expected from DM annihilation. The morphological differences between the two scenarios are investigated. Tests of the impact of different magnetic fields, DM masses, and DM profiles will be performed in the next steps.}

\ConferenceLogo{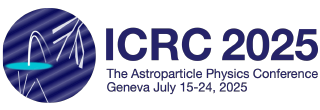}

\FullConference{39th International Cosmic Ray Conference (ICRC2025)\\
 15–24 July 2025\\
Geneva, Switzerland\\}

\newcommand{\dd}{\mathrm{d}}

\begin{document}
\maketitle

\section{Indirect Detection of Dark Matter}

While dark matter (DM) proves to be ever elusive, recent indirect detection attempts in the gamma and radio domains have yielded stringent upper limits on Weakly Interacting Massive Particle (WIMP) DM cross sections.

In the WIMP scenario, DM particles annihilate and produce annihilation products through a variety of annihilation channels. Channels of interest for indirect detection include the bottom quark: $b\bar{b}$, 
top quark: $t\bar{t}$, and W bosons: $W^{+}W^{-}$. While gamma rays are the conventional messenger for DM indirect detection \cite{MAGIC, CTA}, charged particles such as electrons and positrons give supplementary secondary emission in the form of synchrotron radiation and inverse Compton scattering. A comprehensive modeling of the
DM signature is imperative for disentangling it from astrophysical signatures.

Dwarf spheroidal galaxies (dSph) are a convenient testing ground for DM, owing to their
low astrophysical background and high mass-to-light ratios \cite{Gaskins}. Although deep observations of dSph galaxies have provided stringent limits \cite{FermiLAT, MAGICdwarfs}, a clear understanding of the expected nature of astrophysical and DM annihilation signals is missing in the event of future detections. We are left questioning whether an observed signal is due to DM or is cosmic-ray induced, as cosmic rays could have survived the end of star formation or come from a low, maintained star formation rate. We attempt to answer this question with a study comparing the parameter space of cosmic-ray-induced and WIMP-induced secondaries.

For the simulations in the study to result in realistic estimates, it is important to quantify the properties of the galaxy. Due to the low astrophysical signal in dwarf galaxies, it can be difficult to determine their magnetic field structure and source luminosity, which are the main inputs to the simulations. As a first step, we consider the hypothetical scenario where both the DM annihilation and astrophysical emission have the same luminosity in a given dwarf galaxy \cite{Basu}. This allows us to purely consider the morphological differences between the two signals with respect to energy and the galactocentric radius. We simulate the secondaries in a galaxy inspired by transition dwarf galaxies which have just recently become quiescent, as dSphs are difficult to detect in radio frequencies.

In this paper, we first describe the input spectra for WIMP annihilation and their relevant branchings (Section \ref{wimp:sec}), followed by the description of the simulation setup (Section \ref{simulation:sec}). We then discuss the normalization of the injection spectra and spatial distribution of the CR electrons (Section \ref{ssec:CRE_inj}) and the DM annihilation products (Section \ref{ssec:DM_inj}). The results are presented in Section \ref{results:sec} followed by conclusions in Section \ref{conclusions:sec}.

\section{Branching Ratios for WIMP annihilation \label{wimp:sec}}

\begin{figure}[H]
    \centering
    \includegraphics[width=0.8\linewidth]{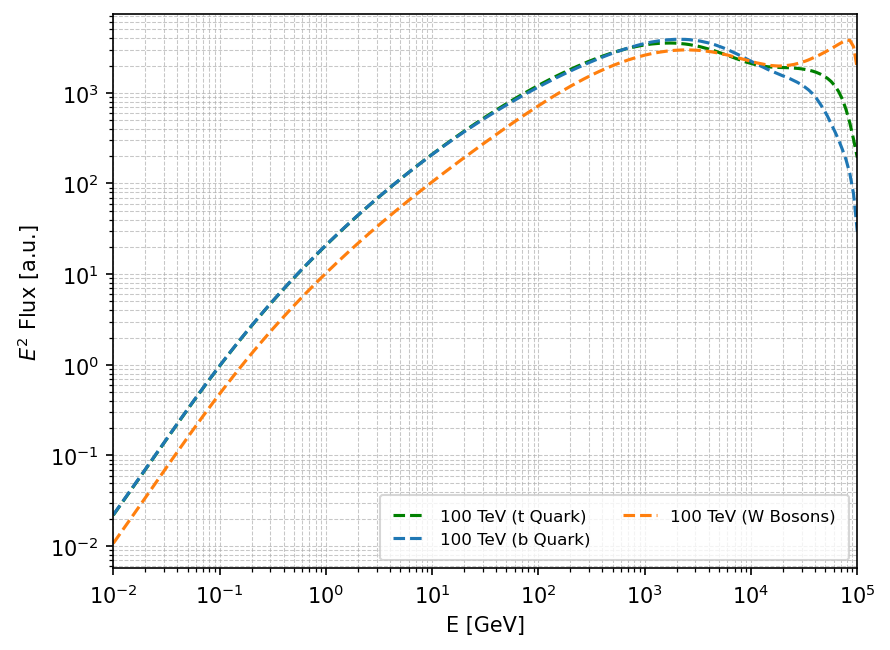}
    \caption{The different annihilation channels of a 100 TeV DM particle assuming 100\% BR into each channel.}
    \label{fig:PPPC_Spectra}
\end{figure}

The DM annihilation spectra are retrieved from the Poor Particle Physicist's Cookbook (PPPC) \cite{PPPC}, which provides the spectra per annihilation event for a given annihilation product. Here, the assumption is that DM annihilates with a 100\% branching ratio (BR) into a given annihilation channel. This allows for consideration of the DM signal for each isolated channel. In this work, the $b\bar{b}$, $t\bar{t}$ and $W^{+}W^{-}$ channels that lead to electrons and positrons for a $10^5 \,\mathrm{GeV}\ (100\,\mathrm{TeV})$ DM particle are considered as they are dominant final states for many weak-scale DM candidates. 
See Figure \ref{fig:PPPC_Spectra}. 

\section{Simulation Setup \label{simulation:sec}}

To simulate the electron content and its secondary emission signals within a dwarf galaxy, we use the publicly available Monte-Carlo framework \texttt{CRPropa} \cite{CRPropa3.2}. \texttt{CRPropa} can transport particles via two different methods, ballistic propagation via the solution of the equation of motion, and diffusive propagation via the solution of the transport equation in the mathematical representation of stochastic differential equations. In this approach, individual phase-space elements are propagated in a particle-like way, allowing the reweighting of the simulation results in the post-processing.

As we expect electrons and positrons in the relevant energy range to diffuse \citep{BeckerTjus&Merten, Ptuskin}, we use the transport equation approach in \texttt{CRPropa}, presented in \cite{CRPropa3.1}. 

In this work, we consider a synthetic dwarf galaxy with maximum radius, $r_\mathrm{m} = 0.5$ kpc. We simulate the electrons with a uniform spatial source distribution. The energy distribution is flat in $\log(E)$, ranging from $10^{-2}$ GeV to $10^5$ GeV. This flat distribution in space and energy allows a re-weighting with equal statistical uncertainty in all space and energy bins. 
We used isotropic and spatially constant diffusion in the dwarf galaxy, applying a Milky Way-like value for the diffusion coefficient with a \textit{Kolmogorov}-like energy scaling.

The radiation loss processes considered are synchrotron radiation and inverse Compton. The magnetic field 
is modeled to be decreasing radially as:
\begin{equation}
    B = \frac{B_0}{1+(\frac{r}{r_0})^2}
    \label{eqn: B_radial}
\end{equation}
where $B_0 = 6 \, \mathrm{\upmu G}$. The background photon field is the Cosmic Microwave Background (CMB).

\section{Normalization of the injection \label{norm:sec}}
In this work, we simulate a generic dwarf galaxy for the two case scenarios: (1) multi-wavelength emission from radiating electrons and positrons, generated via WIMP annihilations; (2) multi-wavelength emission from radiating astrophysical cosmic-ray electrons. The aim is to compare the signatures at equal luminosities for the two cases and to identify regions in the parameter space that are dominated by either cosmic rays or DM signatures to prepare for possible detections of dark dwarfs and the interpretation of the result.

\texttt{CRPropa} conveniently allows for the reweighting of source parameters such as the energy spectrum and spatial distribution during post-processing. By using \texttt{CRPropa} for the propagation of DM secondaries as well as the simulated dwarf galaxy's cosmic rays, the astrophysical and DM emission is treated in a self-consistent framework. 

For the purpose of equal statistics across energy and spatial volume bins, the simulated source distribution is as follows:  

\begin{equation}
    \left. \frac{\dd N}{\dd E \, \dd V}\right|_\mathrm{source} = \frac{N_\mathrm{sim}}{\ln (E_\mathrm{max} / E_\mathrm{min}) \, V} E^{-1}, 
\end{equation}
where $N_\mathrm{sim} = 1 \times 10^8$ is the number of simulated (pseudo-) particles, $E_\mathrm{max} = 10^5 \,\mathrm{GeV}\ (100\,\mathrm{TeV})$ and $E_\mathrm{min} = 10^{-2} \,\mathrm{GeV}\ (10 \,\mathrm{MeV})$ are the maximum and minimum simulated energy, and $V = 4\pi / 3 \, r_\mathrm{m}^3$ is the simulation volume with maximum radius $r_\mathrm{m} = 0.5 \, \mathrm{kpc}$. 

For the desired source scenario $\dd N / \dd E /\dd V|_\mathrm{real}$, one has to apply a weight 
\begin{equation}
    w_i = \frac{ \left. \frac{\dd N}{\dd E \dd V} \right|_\mathrm{real}\left( E_{0,i} , \vec{r}_{0,i}\right)}{\left. \frac{\dd N}{\dd E \dd V}\right|_\mathrm{source}\left( E_{0,i} , \vec{r}_{0,i}\right)}  
\end{equation}
to the $i$-th simulated pseudo-particle with an initial energy $E_{0,i}$ and position $\vec{r}_{0,i}$. 
The weights are calculated for both the CR electron injection and the WIMP annihilation. 

To have a comparable case between the injection scenarios for DM induced electrons and astrophysical cosmic-ray electrons, we choose the absolute normalization of the source term $N_0$ to result in the same total injected energy
\begin{equation}
 E_\mathrm{tot} = \int
\limits_{E_\mathrm{min}}^{E_\mathrm{max}} 
    \dd E \, \int\limits_{0}^{r_\mathrm{m}} \dd r \, 4\pi r^2 E\,  \frac{\dd N}{\dd E \, \dd V} \quad .
\end{equation}
The source normalization follows as,
\begin{equation}
    N_0^\mathrm{astro} = \frac{E_\mathrm{tot}}{\ln(E_\mathrm{max}/E_\mathrm{min})} \, \frac{3(r_\mathrm{m}^2 + r_0^2)^{3/2}}{4\pi \, r_0^3 r_\mathrm{m}^3} \quad.
\end{equation}

\subsection{CR electron injection} \label{ssec:CRE_inj}

For the astrophysical injection scenario, one can consider a power-law source spectrum following $E^{-\alpha}$ where the spectral index, $\alpha$, is 2. The spatial distribution of the sources is defined by the Plummer profile \cite{Plummer}. 
Thus, the source term 
is given by,
\begin{equation}
    \left. \frac{\dd N}{\dd E \dd V} \right|_\mathrm{astro} (E, \vec{r}) = N_0 \, \left(\frac{E}{E_0} \right)^{-2} \, \left(\frac{1}{1 + (r/r_0)^2}\right)^{5/2} \quad , 
\end{equation}
where $E_0 = 1\, \mathrm{GeV}$ is the normalization energy and $r_0 = 0.1\, \mathrm{kpc}$ is the radial scale.

\subsection{WIMP annihilation electron injection} \label{ssec:DM_inj}
In the case of the DM induced injection of electrons and positrons from WIMP annihilation, the spectral energy distribution $\dd N/\dd E$ is calculated from the PPPC (see sec.~\ref{wimp:sec}) and the spatial distribution scales with the square of the DM distribution $\rho_{\rm DM}$. The injection term reads 
\begin{equation}
    \left. \frac{\dd N}{\dd E \dd V}\right|_\mathrm{DM} = N_0 \, \frac{\dd N}{\dd E} \rho_{\rm DM}^2. 
    \label{eq:DMinj}
\end{equation}
In this work, the DM distribution for the NFW \eqref{eqn: NFW} and the Einasto \eqref{eqn: Einasto} 
\begin{subequations}\label{eqn: DM_profiles} are used:
\begin{align}
\rho_{\rm DM}^{\rm NFW}(r) = ~\rho_0\,\left[\dfrac{1}{(r/r_0)\,(1+r/r_0)^2}\right], \label{eqn: NFW}
\\
\rho_{\rm DM}^{\rm Einasto}(r) = ~\rho_0\,{\rm exp}\left[-4\,\left\{\left(\dfrac{r}{r_0}\right)^{1/2} - 1\right\}\right]. \label{eqn: Einasto}
\end{align} 
\end{subequations}

Analogously to the CR electron injection, we fix the normalization of the source distribution by the total injected energy $E_\mathrm{tot}$. Therefore, we introduce the average emitted energy per DM annihilation
\begin{equation}
    \tilde{E} = \int\limits_{E_\mathrm{min}}^{E_\mathrm{max}} \,E \frac{\dd N}{\dd E} \, \dd E \quad ,
\end{equation}
and calculate the normalization constant in eq.~\eqref{eq:DMinj}.
For the NFW profile it reads as 
\begin{equation}
    N_0^\mathrm{NFW} = \frac{12\pi \, E_\mathrm{tot}}{\tilde{E} \rho_0^2}   \left(\frac{1}{8r_0^3}- \frac{r_0^6}{(r_0 + r_\mathrm{m})^3}\right)^{-1} \quad.
\end{equation}
In the case of the Einasto profile, we only consider the solution where the total extent of the galaxy is large compared to the scale radius $r_\mathrm{m} \gg r_0$. The normalization then reads as: 
\begin{equation}
        N_0^\mathrm{Einasto} = \frac{4\pi \, E_\mathrm{tot}}{\tilde{E} \rho_0^2}\frac{16384}{15} \mathrm{e}^{-8} \, r_0^{-3} \quad .
\end{equation}


\section{Results \label{results:sec}}

\begin{figure}[htbp]
    \centering
    \includegraphics[width=\linewidth]{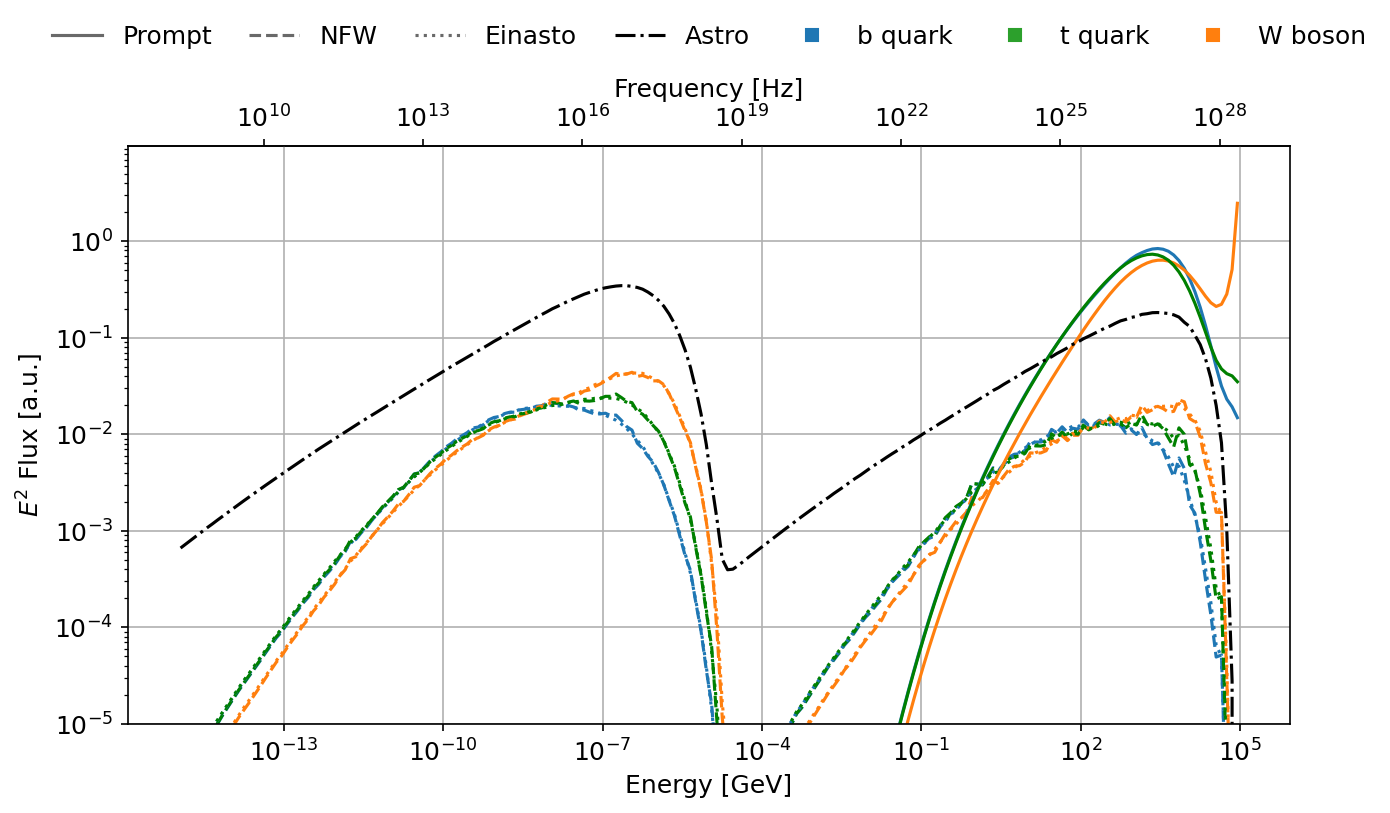}
    \caption{The SED for a dwarf galaxy consisting of a 100 TeV DM particle annihilating into electrons/positrons and astrophysical sources of electrons/positrons. The NFW (dashed) and Einasto (dotted) lines depict the DM secondary emission, and prompt (solid) lines depict the prompt gamma photons assuming 100\% BR into each of the three channels: b quark (blue), t quark (green), and W boson (orange). The black dotted-dashed line shows the astrophysical emission.}
    \label{fig:SED}
\end{figure}

We used the annihilation spectra for a 100 TeV DM particle for the three channels as shown in Figure \ref{fig:PPPC_Spectra} as injection spectra into \texttt{CRPropa} as described above. In this way, we create secondary synchrotron and inverse Compton photons for NFW and Einasto spatial distributions in addition to the prompt gamma photons. The results can be seen in Figure \ref{fig:SED}, where the DM signatures (colored lines, described in detail in the caption of the figure) are shown together with the astrophysical emission (dotted-dashed line in black). In the assumption of 100\% BR to the respective DM channels, the DM secondary emission rises and falls more steeply than the smoother astrophysical spectrum, resulting in sharper peaks in the SED. The choice of the annihilation channel has more impact on the secondary DM spectra than the choice of the spatial profile; the b quark channel yields a softer, lower bump at higher energies while the t quark and W boson channels have harder spectra with slightly more pronounced bumps.

\begin{figure}[htb]
    \centering
    \includegraphics[width=0.8\linewidth]{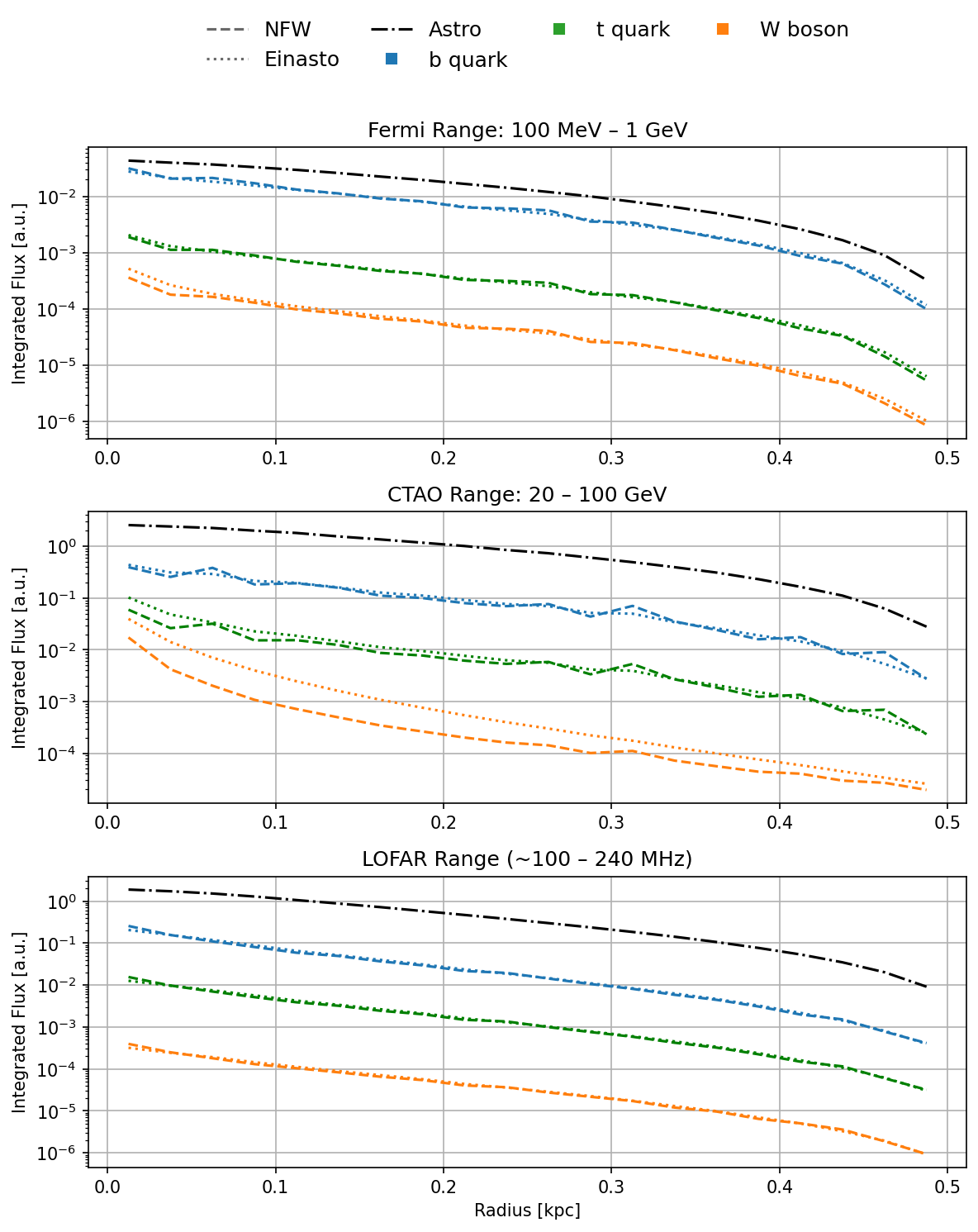}
    \caption{Radial profiles of DM and astrophysical emission across three energy bands. The prompt gamma and secondary emission for the three annihilation channels are summed and plotted together. Each channel considers the two spatial distributions: NFW (dashed line) and Einasto (dotted line). The astrophysical profile is again given by the dash-dotted black line. \textbf{Top:} Range of 100 MeV to 1 GeV, comparable with Fermi-LAT. \textbf{Middle:} Range of 20 to 100 GeV, comparable with the lower end of CTAO. \textbf{Bottom:} Range of ~100 to 240 MHz, comparable with LOFAR frequencies. }
    \label{fig:Radial}
\end{figure}

For the radial profiles in Figure \ref{fig:Radial}, the prompt DM emission into gamma-rays and the secondary DM emission from electron primaries are summed and plotted together for each channel and for both spatial distributions. In the top of Figure \ref{fig:Radial}, we start with the energy range of 100 MeV to 1 GeV, corresponding to the approximate energy range of the Fermi Large Area Telescope (Fermi-LAT). We see a rise near the center for the DM profiles and a more rapid decline with radius than the astrophysical profile, with the exception of the b quark DM profile which is parallel to the astrophysical one. 

For the range of ~20-100 GeV (middle of Figure \ref{fig:Radial}), compatible with the lower end of the Cherenkov Telescope Array Observatory (CTAO), the prompt DM profiles are also higher in the center and decline more rapidly with radius compared to the astrophysical profile. Here, the effect is more prominent than in the Fermi range, heightening the possibility of distinguishing between the two profiles.

For the radio range of $\sim 100$ to 240 MHz (bottom of Figure \ref{fig:Radial}), compatible with the LOw Frequency ARray (LOFAR), the DM and astrophysical profiles are more similar in shape. This could suggest that higher energy gamma ray bands might be a more convenient window to disentangle morphological differences.

\FloatBarrier
\section{Conclusions and Outlook \label{conclusions:sec}}
In this work, we present the study of a generic dwarf galaxy with the assumption of equal luminosities in a WIMP signal and a cosmic-ray electron signal in the multi-wavelength spectrum. We propagate the particles in the galaxy and calculate their emissions from synchrotron radiation
and Inverse Compton scattering. We evaluate the parameter space of the emissions as a function of
the galactocentric radius and energy.

 When considering the radial profiles in the higher energy gamma ray bands, the DM profiles for both spatial distributions are more concentrated in the centre of the simulated galaxy and fall off more rapidly than the astrophysical profile.

One point of discussion is the ratio of the strength of the prompt gamma emission and the secondary radio emission, as the latter is susceptible to change with the magnetic field strength. Furthermore, the choice of the DM mass will also influence what becomes visible in each of the energy windows of the radial profiles. This motivates future steps, which include different field morphologies and DM masses. 

The consideration of a time-varying magnetic field can also be of significance as dwarf galaxies with no active star formation phase will have their magnetic fields decaying over time. A time-varying magnetic field will have to be taken into account for a more accurate description of charged particle propagation in dwarf galaxies. 

{\small 
\section*{Acknowledgments}
\noindent
AA, JD, JBT, DB, SD, ST and DE acknowledge funding from the German Science Foundation DFG, via the Collaborative Research Center SFB1491 - "Cosmic Interacting Matters - From Source to Signal" (project no.\ 445052434).
}

\setlength{\bibsep}{0pt plus 0.3ex}


\begin{thebibliography}{99}






\bibitem{MAGIC}
MAGIC Collaboration, S. Abe et al., \doilink{Journal of Cosmology and Astroparticle Physics}{10.1088/1475-7516/2025/03/020} (2025)

\bibitem{CTA}
CTA Collaboration, S. Abe, J. Lazendic‑Galloway, et al., \doilink{Journal of Cosmology and Astroparticle Physics}{10.1088/1475-7516/2024/07/047} (2024)

\bibitem{Gaskins}
J. M. Gaskins,
\doilink{Contemporary Physics 57}{https://doi.org/10.1080/00107514.2016.1175160} (2016).

\bibitem{FermiLAT}
A. McDaniel, M. Ajello, C. M. Karwin, et al., 
\doilink{Physical Review D}{https://doi.org/10.1103/PhysRevD.109.063024} (2024).

\bibitem{MAGICdwarfs}
MAGIC Collaboration, V. A. Acciari et al., 
\doilink{Physics of the Dark Universe}{https://doi.org/10.1016/j.dark.2021.100912} (2022).

\bibitem{Basu}
A.Basu, A. Aravinthan, J. Becker Tjus, et  al., {in preparation}.

\bibitem{PPPC}
M. Cirelli, G. Corcella, A. Hektor,
\doilink{Journal of Cosmology and Astroparticle Physics}{https://doi.org/10.1088/1475-7516/2011/03/051}(2011)

\bibitem{CRPropa3.2} R. Alves Batista, J. Becker Tjus, J. D\"orner, et al. \doilink{Journal of Cosmology and Astroparticle Physics}{10.1088/1475-7516/2022/09/035} (2022)

\bibitem{CRPropa3.1} L. Merten, J. Becker Tjus, H. Fichtner, et al. \doilink{Journal of Cosmology and Astroparticle Physics}{10.1088/1475-7516/2017/06/046} (2017)

\bibitem{Plummer}
H.C. Plummer, \doilink{Monthly Notices of the Royal Astronomical Society 71}{10.1093/mnras/71.5.460} (1911) 

\bibitem{BeckerTjus&Merten}
J. Becker Tjus, L. Merten, \doilink{Physics Reports 872}{10.1016/j.physrep.2020.05.002} (2020) 


\bibitem{Ptuskin}
V. S. Ptuskin, \newlink{ICRC 2005}{https://cds.cern.ch/record/965856/files/20317_Vladimir_Ptuskin_060210.pdf} (2005)



\end{thebibliography}
\end{document}